\documentstyle[12pt,epsfig,amsmath,amssymb,amsbsy]{article}
\oddsidemargin   - 1.1cm
\def\carbon{$^{13}C$}
\def\proton{$^1H$}
\def\spf{$(TMTTF)_2PF_6$}

\def\sasf{$(TMTTF)_2AsF_6$}
\def\ssbf{$(TMTTF)_2SbF_6$}
\def\CiT1{$^{13}T_1^{-1}$}
\def\PiT1{$^{1}T_1^{-1}$}
\def\a{${\mathbf{a}}$}

\begin{document}
\pagestyle{empty}

\fontfamily{ptm}
\selectfont
\noindent
{\Large \bf Influence of Charge Order on the Ground States of TMTTF Molecular Salts}
\vspace{1cm}

\noindent
W. Yu$^1$, F. Zamborszky$^{1,2}$, B. Alavi$^1$, A. Baur$^3$, C. A. Merlic$^3$, and S. E. Brown$^1$

\vspace{.2cm}
\noindent
$^1${\it Department of Physics, UCLA, Los Angeles, CA 90095 USA}\\
$^2${\it Los Alamos National Laboratory, Los Alamos, NM 87545 USA}\\
$^3${\it Department of Chemistry and Biochemistry, UCLA, Los Angeles, CA 90095 USA}\\
\vspace{1cm}

\noindent
\hfill  \parbox{16.5cm}{
\footnotesize {\bf Abstract.}
\sasf\ and \ssbf\ are both known to undergo a charge ordering phase transition, though their ground states are 
different. The ground state of the first is Spin-Peierls, and the second is an antiferromagnet. We study the effect of 
pressure on the ground states and the charge-ordering using \carbon\ NMR spectroscopy. The experiments demonstrate 
that the the CO and SP order parameters are repulsive, and consequently the AF state is stabilized when the CO order 
parameter is large, as it is for \ssbf. An extension of the well-known temperature/pressure phase diagram is proposed.
\vspace{.5cm}

}

\vspace{.9cm}

\noindent
{\bf 1. INTRODUCTION}
\vspace{.4cm}

\noindent
Unlike the highly-conducting TMTSF analog compounds, molecular salts based on $TMTTF$ molecules tend to insulating 
behavior below $T=100-300K$. It is commonly attributed to a dimerization of the $TMTTF$ stacks that results in a 
half-filled anti-bonding band $[1]$. A Mott-Hubbard gap results, without additional symmetry breaking, once electronic 
correlations are included. If the dimerization is sufficiently weak compared to the interchain hopping integrals, then 
delocalization occurs. Charge-ordering (CO) is an alternative route to an insulating state for 1/4-filled systems. 
Near-neighbor repulsive Coulomb interactions tend to stabilize a CO phase $[2]$, but lattice degrees of freedom can 
also be very important $[3,4]$. 

Recently, \carbon\ NMR spectroscopy in salts with $AsF_6$ and $PF_6$ counterions demonstrated a phase transition to a 
CO state $[5]$ in both systems. It was shown that two inequivalent molecular environments develop below what appeared 
to be a continuous phase transition. Each occurs in equal numbers; the natural expectation for the order parameter is 
CO. Low-frequency dielectric measurements $[6]$ yield an anomalously large polarizibility following a Curie-Weiss Law, 
consistent with the breaking of an inversion symmetry of the unit cell. Following this hypothesis, the charge 
configuration along the stacks is expected to be in a $'1010'$ pattern, with $1(0)$ symbolizing charge rich (poor) 
sites, and each stack ordering equivalently. 

In addition to the compounds with the centrosymmetric counterions $AsF_6$, $PF_6$, and $SbF_6$, salts with 
non-centrosymmetric anions such as $ReO_4$ or $SCN$ undergo a CO transition that is either coincident with anion 
ordering ($SCN$) or not ($ReO_4$) $[7,8,9]$. Given that the phenomenon is commonplace, it is important to establish 
whether the interactions producing the CO, as well as the CO itself, play any significant role in controlling the 
ground state. There are two levels to which this question should be asked. The first is relevant to the $TMTTF$ salts, 
which are all insulators at intermediate temperatures in the sense that the resistivity along the stack direction is 
increasing upon cooling below temperatures of order $100-200K$. Restricting the discussion to the case of 
centrosymmetric counterions, the ground state is either Spin-Peierls ($AsF_6$, $PF_6$), or antiferromagnetic ($SbF_6$, 
$Br$). The next logical step is to ask what role charge fluctuations, resulting from the interactions that produce CO 
in the insulators, play in the highly conducting (and superconducting) $TMTSF$ salts $[10,11,12]$. Here we focus on 
the first of these questions, starting by reviewing the results of high-pressure experiments on the \sasf\ salt. These 
demonstrate a repulsive coupling of the Spin-Peierls (SP) and CO order parameters. A description of \carbon\ NMR 
experiments on the $SbF_6$ salt follows. We learn from those experiments that for sufficiently strong CO, the SP 
ground state is suppressed and an antiferromagnetic (AF) state is stabilized in its place. Application of pressure 
reduces the amplitude of the CO and restores the SP state. The experiments enable us to properly map the $SbF_6$ salt 
onto the well-known temperature/pressure (T/P) phase diagram for the $TMTTF$ and $TMTSF$ compounds $[13]$.

\vspace{.6cm}

\noindent
{\bf 2. EXPERIMENTAL} 
\vspace{.4cm}

\noindent
\carbon\ spin-labeled $TMTTF$ molecules were synthesized at UCLA $[14]$, and the $(TMTTF)_2X$ crystals were grown by 
the standard electrolysis method. Hydrostatic pressure was applied using a standard BeCu clamp cell with FC-75 (3M) 
serving as the pressure medium. Pressures are inferred from the forces applied at $T=300K$. All experiments presented 
here were performed in an applied field of $B_0=9T$, with the molecular stacking (\a) axis perpendicular to the field. 

\vspace{.4cm}

\noindent
{\bf 2.1 Results for \sasf}

\vspace{.4cm}

\begin{figure}[ht]
\includegraphics[width=8cm]{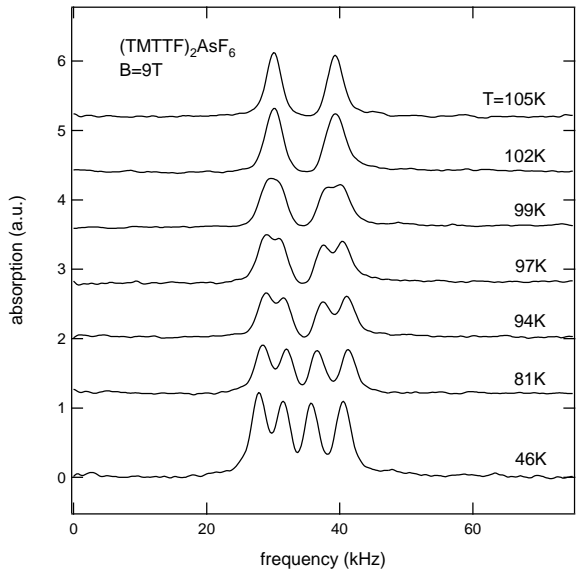}
\caption{\footnotesize{\carbon\ NMR spectra at temperatures $T<T_{CO}$ and $T>T_{CO}$. The applied field is 
$B_0=9.0T$, directed in the $\mathbf{b'}-\mathbf{c*}$ plane approximately $55^\circ$ from the molecular symmetry axis 
$[5]$.}}
\label{sasfSpectrum} 
\end{figure}

\noindent
The experimental signature of the CO phenomenon using \carbon\ NMR spectroscopy appears in Fig. \ref{sasfSpectrum} for 
\sasf, recorded at temperatures on both sides of the transition at $T_{CO}=103K$ $[5]$. At high temperatures, each 
molecule is equivalent; the two NMR lines correspond to the inequivalent sites located on the bridge of the dimer 
molecule. The distinction could be defined in terms of their locations relative to the nearest counterion. On cooling 
through $T_{CO}$, the two lines split into four. Independent experiments verify that the four lines originate with 
nuclei in two different molecular environments. 

\begin{figure}[htb]
\includegraphics[width=8cm]{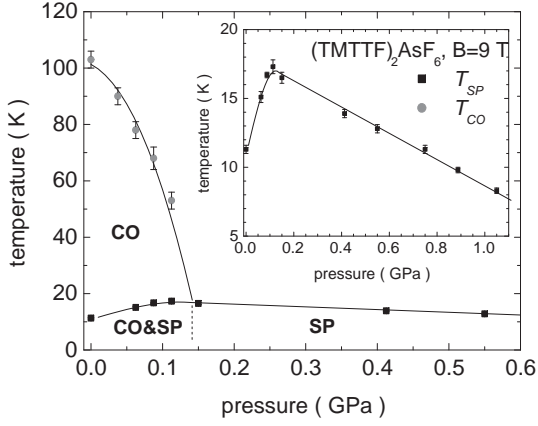}
\caption{\footnotesize{Phase diagram of \sasf\ established from \carbon\ NMR experiments. The solid lines are a guide 
to the eye, and the dashed line is used only to emphasize that there is a region of coexistence (see text).}}
\label{sasfpd} 
\end{figure}

In Fig. \ref{sasfpd} are the results of a high pressure study on the stability of the CO phase in \sasf. $T_{CO}$ is 
suppressed rapidly with applied pressure, and there is a striking variation of the critical temperature of the SP 
ground state $T_{SP}$. At low pressures, we have $dT_{CO}/dP<0$ and $dT_{SP}/dP>0$. At high pressures, where there is 
no evidence for a distinct CO transition, $dT_{SP}/dP<0$ as it has been reported for \spf $[15]$. We take the 
experimentally established phase diagram as evidence for a repulsive coupling of the CO and SP order parameters 
$[16]$. 

These observations led us to consider the antiferromagnetic system \ssbf. In particular, we wanted to understand 
whether the AF state was the favored ground state when the CO amplitude is large. The high-pressure NMR study, 
described below, shows that it is. When the pressure is increased sufficiently to weaken the CO by reducing $T_{CO}$ 
as well as the amplitdue of the charge disproportionation, a singlet ground state results. Most likely, it is an SP 
state. We propose an extension of the T/P phase diagram for the $TMTTF/TMTSF$ family that includes the possibility for 
a reentrant AF phase at $T=0$.    

\vspace{.4cm}

\noindent
{\bf 2.2 Results for \ssbf}

\vspace{.4cm}

\begin{figure}[htb]
\includegraphics[width=14cm]{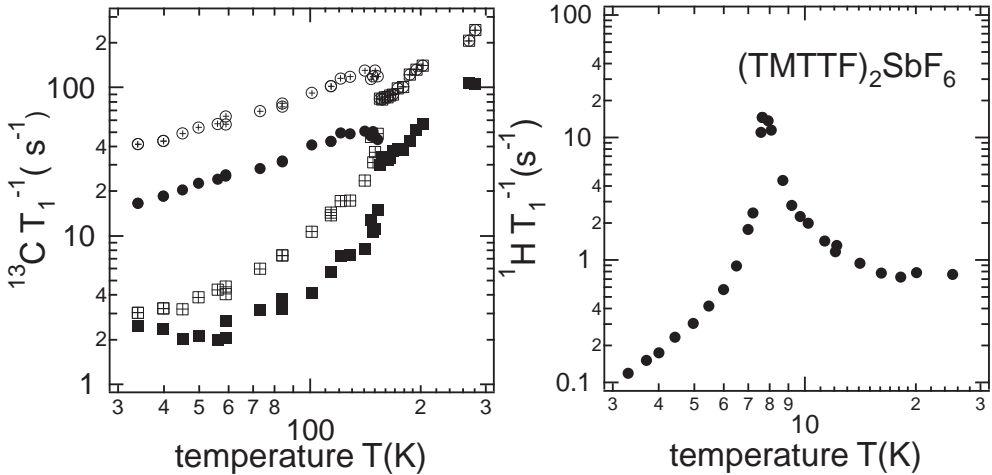}
\caption{\footnotesize {a) \carbon\ spin lattice relaxation rate \CiT1\ vs. temperature. $T_{CO}$ is the temperature 
where the number of inequivalent sites increases from two to four. b) \proton\ spin lattice relaxation rate \PiT1\ vs. 
temperature. The sharp peak marks the antiferromagnetic ordering at $T_N$.}}
\label{iT1SbF6} 
\end{figure}

\noindent
In Fig. \ref{iT1SbF6}a we show the temperature dependence of the \carbon\ spin-lattice relaxation rate. For 
$T>T_{CO}=156K$, there are two rates for the inequivalent sites of each molecule. Reducing the temperature through 
$T_{CO}$ leads to two inequivalent molecular environments (A, B), so there are four distinct rates. If the local site 
occupancy is $n_A$ and $n_B$, we can obtain an estimate of the charge disproportionation $\Delta 
n=|n_A-n_B|/(n_A+n_B)$ by assuming that 

\begin{equation}
\frac{T_{1A}^{-1}}{T_{1B}^{-1}}=\frac{n_A^2}{n_B^2},
\end{equation}

\noindent
and therefore $\Delta n\approx .5$ for \ssbf\ at low temperatures. The ratio of relaxation rates is only slightly 
larger than what we observed for \sasf $[15]$. Raman studies of the intramolecular vibrational frequencies gave a 
smaller value for the disproportionation in \sasf, $\Delta n=0.34$ $[17]$. The transition to the AF state is clearly 
identified as a peak in the $^1H$ spin lattice relaxation rate at $T_N(B=9T)\approx 7K$, as shown in Fig. 
\ref{iT1SbF6}b. 

\begin{figure}
\includegraphics[width=14cm]{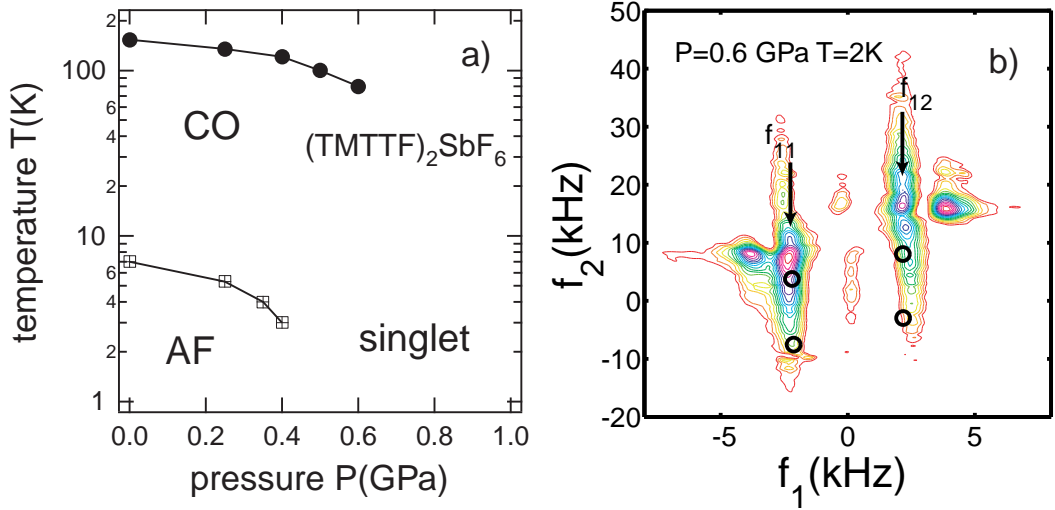}
\caption{\footnotesize{ a) Observed variation of $T_{CO}$ and $T_N$ with pressure in \ssbf. b) 2D \carbon\ NMR 
spectrum at $P=0.6GPa$ and $T=2K$. The open circles mark the location of the four features in the spectrum at 
$T\approx 30K$.}}
\label{2DHiPloT} 
\end{figure}

In Fig. \ref{2DHiPloT}a, the evolution of the CO and AF phase transitions with applied pressure is shown. We note that 
$T_N$ decreases with applied pressure, along with $T_{CO}$. When the pressure exceeds about $0.5GPa$, the NMR 
signatures for the CO state are nearly nonexistent, even though the ordering temperature is suppressed to only about 
$(1/2)T_{CO}(P=0)$. Transport measurements $[18]$ also indicate that the signatures for the CO state diminish rapidly 
with pressure. No evidence for the AF state is seen at high pressures. 

Instead, another ground state was identified by using two-dimensional (2D) NMR techniques. First, the sample is 
rotated away from the magic angle orientation, so that the internuclear dipolar interaction is nonzero. In a 1D 
experiment, the number of peaks doubles relative to what is shown in Fig. \ref{sasfSpectrum}. The data set is then 
constructed from a standard echo experiment: for each spin echo transient recorded over the interval ($t_2$), there is 
a pulse separation time $t_1/2$. The Discrete Fourier Transform with respect to $t_2$ and $t_1$ are the frequencies 
$f_2$ and $f_1$. The internuclear coupling leads to a modulated echo decay corresponding to the peak separation along 
the $f_1$ dimension. The hyperfine coupling {\it and} the dipolar coupling produce shifts along the $f_2$ dimension. 
The results are shown in Fig. \ref{2DHiPloT}b. In the high-symmetry phase are the expected four peaks, located at two 
frequencies ($f_{11}$ and $f_{12}$) in the $f_1$ dimension. The four open circles mark the position of these four 
peaks at $T=30K$ and $P=0.5GPa$. The contours shown are from data recorded at $T=2K$ and the same pressure, $0.5GPa$. 
There is a spread of the spectrum in the $f_2$ dimension at the same $f_1$ frequencies as at high temperature. Also 
seen are two sharp features at new $f_1$ frequencies. 

\vspace{.4cm}

\noindent
{\bf 3. DISCUSSION}

\vspace{.4cm}

\noindent
The pair of peaks shifted from $f_{11}$ and $f_{12}$ to $(3/2)f_{11}$ and $(3/2)f_{12}$ is the spectrum of coupled, 
equivalent sites, and correspond to \carbon\ pairs with negligible hyperfine shifts. A spectrum like this is 
characteristic of the singlet Spin-Peierls phase $[5]$. Following this interpretation, the portion of the spectrum 
broadened along $f_2$, but remaining at $f_{11}$ and $f_{12}$, result from the creation of domain walls. That is, 
$B=9.0T>B_c$, with $B_c$ the critical field for triplet excitations $[19]$. 

\begin{figure}
\includegraphics[width=8cm]{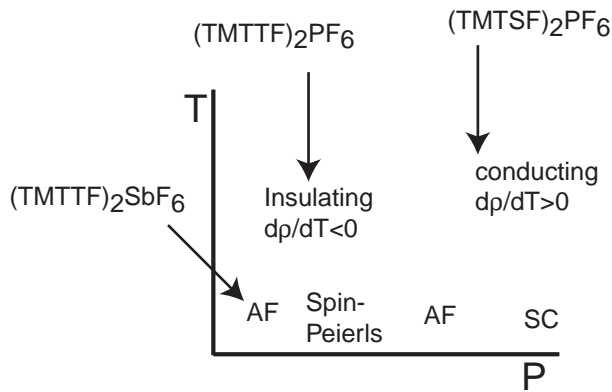}
\caption{\footnotesize{An extended version of a schematic temperature/pressure phase diagram for the $TMTTF$ and 
$TMTSF$ salts.}}
\label{TMTCFphases} 
\end{figure}

Based on the experimental observations, we conclude that not only does the CO tend to suppress the SP order, but that 
the ambient pressure AF state is favored when the CO is particularly stable. As soon as the CO is sufficiently 
weakened through the use of hydrostatic pressure, the SP state is reestablished. The principle consequence of our 
experiments is a new understanding of how the CO impacts the evolution of the ground state by pressure, and also how 
the \ssbf\ compound fits into a T/P phase diagram with the other salts made using centrosymmetric counterions. Our 
version of a schematic representation is shown in Fig. \ref{TMTCFphases}. At high temperatures, the identifiable 
difference between materials is whether they tend to be insulating ($d\rho/dT<0$) or metallic ($d\rho/dT>0$). The 
antiferromagnetic state of \ssbf\ is suppressed with applied pressure and it becomes more like \sasf\. Presumably, 
with enough pressure it could be made to superconduct. 

In summary, we have presented results from high-pressure NMR studies of two TMTTF salts with centrosymmetric 
counterions. The ground states for the two salts are different: \sasf\ has a Spin-Peierls ground state, and the ground 
state of \ssbf\ is AF. By studying the effect of high pressure on the physical properties of these two materials, we 
have shown that the CO and SP order parameters are repulsive, and the AF ground state of the \ssbf\ material is a 
natural consequence of the CO phenomenon. The well-known temperature/pressure diagram is revised to accommodate the 
findings.

\vspace{.5cm}

\vspace{.4cm}

\noindent
{\bf Acknowledgements}

\vspace{.4cm}

\noindent
The research was supported by the National Science Foundation under grant number DMR-0203806. The authors are grateful 
for conversations with S. Brazovskii, M. Dressel, M. Dumm, H. Fukuyama, P. Monceau, H. Seo, and M. Ogata.

\vspace{.4cm}

\noindent
{\bf References}
\vspace{0.4cm}

\noindent
$[1]$ V. J. Emery, R. Bruinsma, and S. Barisic, {\it Phys. Rev. Lett.} {\bf 48}, 1039-1043 (1982).\\
$[2]$ H. Seo and H. Fukuyama, {\it J. Phys. Soc. Jpn.} {\bf 66}, 1249 (1997).\\
$[3]$ S. Mazumdar, R.T. Clay, and D.K. Campbell, {\it Phys. Rev. B} {\bf 62}, 13400, (2000).\\
$[4]$ R. T. Clay, S. Mazumdar, and D. K. Campbell, {\it Phys. Rev. B} {\bf 67}, 115121 (2003).\\
$[5]$ D. S. Chow, {\it et al.}, {\it Phys. Rev. Lett.} {\bf 85}, 1698 (2000).\\
$[6]$ P. Monceau, F. Ya. Nad, and S. Brazovskii, {\it Phys. Rev. Lett.} {\bf 86}, 4080 (2001).\\
$[7]$ C. Coulon, S. S. P. Parkin, and R. Laversanne, {\it Phys. Rev. B} {\bf 31}, 3583 (1985).\\
$[8]$ H. H. S. Javadi, R. Laversanne, and A. J. Epstein, {\it Phys. Rev. B} {\bf 37}, 4280 (1988).\\
$[9]$ S. Brazovskii, {\it J. Phys. IV} {\bf 12} 149 (2002).\\
$[10]$ S. Mazumdar, {\it et al.}, {\it Phys. Rev. Lett.} {\bf 82}, 1522 (1999).\\ 
$[11]$ S. E. Brown, {\it et al.}, {\it Synth. Metals} {\bf 137}, 1299 (2003).\\
$[12]$ S. Brazovskii, P. Monceau, F. Nad, {\it Synth. Metals} {\bf 137}, 1331 (2003).\\
$[13]$ H. Wilhelm, {\it et al.}, {\it Eur. Phys. J. B} {\bf 21} 175 (2001), and references therein.\\
$[14]$ C. A. Merlic, {\it et al.}, {\it Synth. Commun.} {\bf 29}, 2953 (1999).\\
$[15]$ D. S. Chow, {\it et al.}, {\it Phys. Rev. Lett.} {\bf 81}, 3984 (1998).\\
$[16]$ F. Zamborszky, {\it et al.}, {\it Phys. Rev. B} {\bf 66} R081103 (2002).\\ 
$[17]$ M. Dumm and M. Dressel (private communication).\\
$[18]$ P. Monceau (private communication).\\
$[19]$ S. E. Brown, {\it et al.}, {\it Phys. Rev. Lett.} {\bf 80}, 5429 (1998).

\end{document}